\documentclass[aps,prb,twocolumn,superscriptaddress,showpacs,showkeys,floatfix]{revtex4}

\usepackage{graphicx,color}
\graphicspath{{figs/}}
\begin{document}
\title{Adsorbate Migration Effects on Continuous and Discontinuous Temperature-Dependent Transitions in the Quality Factors of Graphene Nanoresonators}
\author{Jin-Wu Jiang}
    \altaffiliation{Electronic address: jwjiang5918@hotmail.com}
    \affiliation{Institute of Structural Mechanics, Bauhaus-University Weimar, Marienstr. 15, D-99423 Weimar, Germany}
\author{Bing-Shen~Wang}
    \affiliation{State Key Laboratory of Semiconductor Superlattice and Microstructure and Institute of Semiconductor, Chinese Academy of Sciences, Beijing 100083, China}
\author{Harold S. Park}
    \altaffiliation{Electronic address: parkhs@bu.edu}
    \affiliation{Department of Mechanical Engineering, Boston University, Boston, Massachusetts 02215, USA}
\author{Timon Rabczuk}
    \altaffiliation{Electronic address: timon.rabczuk@uni-weimar.de}
    \affiliation{Institute of Structural Mechanics, Bauhaus-University Weimar, Marienstr. 15, D-99423 Weimar, Germany}
    \affiliation{School of Civil, Environmental and Architectural Engineering, Korea University, Seoul, South Korea }
%\date{22 December 2009}
\date{\today}
\begin{abstract}
We perform classical molecular dynamics simulation to investigate the mechanisms underpinning the unresolved, experimentally-observed temperature-dependent scaling transition in the quality factors of graphene nanomechanical resonators (GNMR). Our simulations reveal that the mechanism underlying this temperature scaling phenomenon is the out-of-plane migration of adsorbates on GNMRs.  Specifically, the migrating adsorbate undergoes frequent collisions with the GNMR, which strongly influences the resulting mechanical oscillation, and thus the quality factors.  We also predict a discontinuous transition in the quality factor at a lower critical temperature, which results from the in-plane migration of the adsorbate.  Overall, our work clearly demonstrates the strong effect of adsorbate migration on the quality factors of GNMRs.
\end{abstract}

\maketitle
%\tableofcontents
%\pagebreak

\section{introduction}
Graphene nanomechanical resonators (GNMR) have drawn intense attention \cite{EkinciKL,ArletJL,EomK} since the first experiments by Bunch \emph{et al.} in 2007.\cite{BunchJS2007sci} GNMRs are a promising candidate for practical applications like mass sensing due to its desirable combination of high stiffness and large surface area.\cite{LeeC2008sci,JiangJW2009young,JiangJW2010young}  For it to be effective for these applications, it is imperative that the GNMRs exhibit a high quality (Q) factor. The Q-factor can be affected by various energy dissipation mechanisms, such as external attachment energy loss,\cite{SeoanezC,KimSY2009apl} intrinsic nonlinear scattering mechanisms,\cite{AtalayaJ} the effective strain mechanism,\cite{JiangJW2012nanotechnology} edge effects,\cite{KimSY2009nl,JiangJW2012jap}, or grain boundary-mediated scattering losses\cite{QiZ2012nns}.  As a result of both the theoretical studies and improvements in experimental studies of GNMRs, their reported Q-factors have increased considerably in recent years.  For instance, Eichler {\it et al.}~\cite{EichlerA} found that the Q-factors of GNMRs can reach values of $10^{5}$.  The temperature-dependence of the Q-factor has also been studied in several experiments. Bunch \emph{et al.} observed a substantial increase in the Q-factor with decreasing temperature, with the Q-factor reaching 9000 at 10~K.~\cite{ZandeAMVD}.  Chen {\it et al.} also found that the Q-factor of GNMR increases with decreasing temperature, and reaches $10^{4}$ at 5 K~\cite{ChenC2009nn}.

A general characteristic in the temperature dependence of the Q-factor is the temperature scaling phenomenon, i.e the Q-factor increases exponentially with two different exponents above or below a critical temperature ($T_{z}$). In other words, there is a continuous transition at $T_{z}$ in the temperature dependence of the Q-factor, where a clear experimental illustration of this temperature-dependent Q-factor transition can be found in Fig. 5 of the work by~\citet{ZandeAMVD} and in Fig. 6 of the work by~\citet{ChenC2009nn}.  The value of the transition temperature $T_{z}$ varies from different experiments, ranging from 50 to 100~K in the GNMR.\cite{ZandeAMVD,ChenC2009nn} However, a theoretical explanation for this phenomenon has not been presented.  Furthermore, in all previous theoretical studies, only a single exponent was obtained for the temperature dependence of the Q-factor in GNMRs or nanotube resonators.\cite{JiangH2004prl,KimSY2009nl,KimSY2009apl,JiangJW2012jap,QiZ2012nns} Therefore, the aim of the present work is to identify the underlying mechanism for this unresolved temperature scaling phenomenon on the Q-factor of GNMRs.

In this paper, we identify the out-of-plane migration of adsorbate atoms as the cause of this unresolved temperature scaling phenomena on the Q-factors of GNMRs via classical molecular dynamics (MD) simulations.  We identify two critical temperatures, the first being that at which the adsorbate starts to migrate in-plane ($T_{xy}$), and the second being that at which the adsorbate begins to migrate out-of- the plane of oscillation ($T_{z}$). The Q-factor undergoes a discontinuous, step-like decrease at the first critical temperature $T_{xy}$ due to the introduction of strong thermal noise resulting from the in-plane migration.  In contrast, a continuous transition in the Q-factor is observed at the second critical temperature $T_{z}$, which corresponds to the temperature scaling phenomenon observed in the experiments.\cite{ZandeAMVD,ChenC2009nn} We also study the effects of adsorbate percentage and mass on the temperature scaling phenomenon.

\section{simulation details}

The smallest translational unit cell in graphene is chosen to contain four carbon atoms and has the size $2.46\times 4.26$~{\AA}. The graphene sample in our simulations has dimensions $(l_{x}, l_{y})$. The length is $l_{x}=n_{x}\times 2.46$~{\AA} and the width is $l_{y}=n_{y}\times 4.26$~{\AA}, where $n_{x}$ and $n_{y}$ are the number of unit cells in the $x$ and $y$ directions. The atoms at the $+x$ and $-x$ ends of the GNMR are fixed, while periodic boundary conditions are applied in the $y$ direction. The interactions of the carbon atoms are described by the Brenner (REBO-II) potential~\cite{brennerJPCM2002}. For the cases where a single adsorbate atom is adsorbed on the GNMR, the interaction between the adsorbate gold atom and the GNMR is modeled by a Lennard-Jones potential with length parameter $\sigma$=2.9943~{\AA} and energy parameter $\epsilon$=0.02936~{eV}~\cite{KimSYnanotechnology}. If more than one adsorbate atom is considered, the interaction between adsorbates is ignored, and thus our simulation mimics a small adsorbate percentage. The standard Newton equations of motion are integrated in time using the velocity Verlet algorithm with a time step of 1 fs.

Our simulations are performed as follows.  First, a Nos\'e-Hoover\cite{Nose,Hoover} thermostat is applied to thermalize the system to a constant temperature within the NVT (i.e. the particles number N, the volume V and the temperature T of the system are constant) ensemble, which is run for 50000 MD steps. The mechanical oscillation of the resonator is then actuated by adding a velocity distribution to the system, which follows the morphology of the first flexural vibrational mode of graphene~\cite{JiangJW2012jap}.  The imposed velocity distribution, or actuation energy, is $\Delta E=\alpha E_{k}^{0}$, where $E_{k}^{0}$ is the total kinetic energy in the GNMR after thermalization but just before its actuation and $\alpha$ is the actuation energy parameter. We have chosen a small value of $\alpha=1.0$ in all simulations here, so that the mechanical oscillation occurs within the linear regime. After the actuation energy is applied, the system is allowed to oscillate freely within the NVE (i.e. the particles number N, the volume V and the energy E of the system are constant) ensemble for typically $10^{6}$ MD steps. The data from the NVE ensemble is used to analyze the mechanical oscillation, energy dissipation, and Q-factors of the GNMR.

\section{results and discussion}

\begin{figure}[htpb]
  \begin{center}
    \scalebox{0.9}[0.9]{\includegraphics[width=8cm]{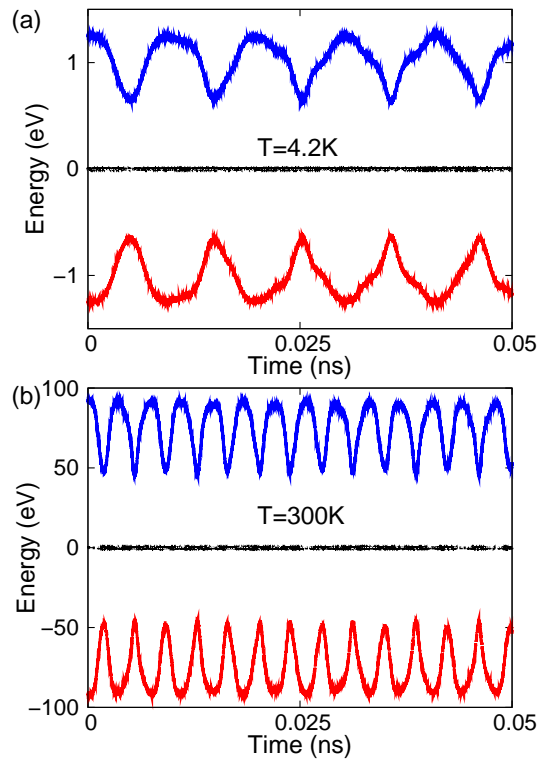}}
  \end{center}
  \caption{(Color online) Energy conservation in the MD simulation for a GNMR of size (50, 6) with a single adsorbate. Kinetic (potential) energy is above (below) the $x$ axis. The potential energy is shifted so that the total energy is zero.}
  \label{fig_energy_conserve}
\end{figure}

\begin{figure}[htpb]
  \begin{center}
    \scalebox{0.9}[0.9]{\includegraphics[width=8cm]{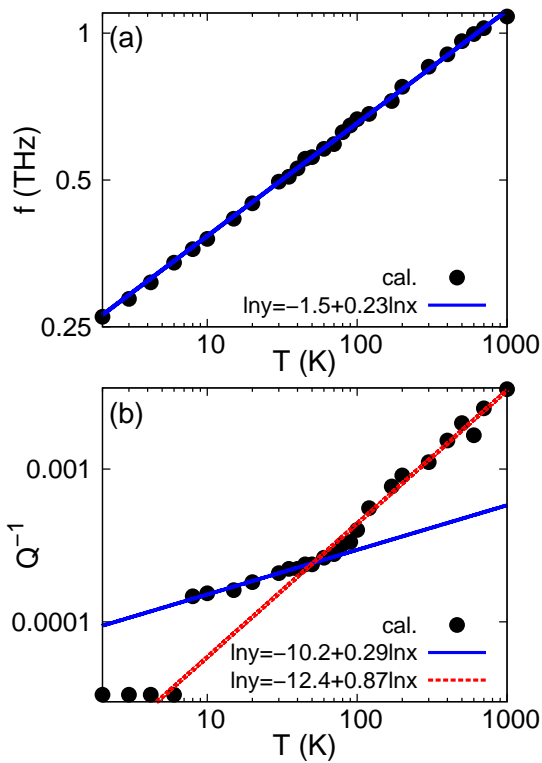}}
  \end{center}
  \caption{(Color Online) The temperature dependence for the frequency and the Q-factor in a GNMR of size (50, 6) with a single adsorbate. (a) shows that the frequency has a clear power law temperature dependence, with power factor 0.23. (b) shows two transition points in the temperature dependence of the inverse of the Q-factor. The first transition at $T_{xy}=7\pm 1$~{K} is discontinuous and is due to the in-plane migration of the adsorbate. The second transition at $T_{z}=47.5\pm 2.5$~{K} is continuous and is caused by the out-of-plane migration of the adsorbate. The second transition corresponds to the temperature scaling observed in experiments.\cite{ZandeAMVD,ChenC2009nn}}
  \label{fig_qfactor}
\end{figure}

Figure.~\ref{fig_energy_conserve} demonstrates that total energy is conserved in the MD simulation within the  NVE ensemble for a GNMR of size (50, 6) with a single adsorbate, where the kinetic (potential) energy is above (below) the $x$ axis. The potential energy is shifted by such a value that the total energy is zero, and the temperatures are 4.2 K and 300 K in these two panels. The mechanical oscillation is actuated at $t=0$ by adding the velocity distribution, giving the maximum kinetic energy at $t=0$. The mechanical oscillation of the GNMR is reflected by the energy exchange between the kinetic and potential energy. We note that the oscillation frequency is clearly higher at 300~{K} than at 4.2~{K}.

Figure.~\ref{fig_qfactor} shows the temperature dependence for the frequency and the Q-factor in the GNMR of size (50, 6) with a single adsorbate, where both plots are presented in log-log scale. The top panel shows that the frequency has a clear power law temperature dependence, with power factor 0.23.  The frequency increases with increasing temperature, which results from the vibration-induced effective strain within the GNMR.\cite{JiangJW2012nanotechnology} The effective strain becomes stronger at higher temperature, resulting in higher oscillation frequencies at higher temperature. A different behavior was observed experimentally,\cite{ZandeAMVD} which may be due to possible electron-phonon interactions from the high electronic current applied in the experiment.

Figure.~\ref{fig_qfactor}~(b) shows two distinct transitions for the temperature dependence of the inverse of the Q-factor.  The first transition is discontinuous and occurs at the temperature $T_{xy}=7\pm 1$~{K}. The second transition is continuous and occurs at $T_{z}=47.5\pm 2.5$~{K}. At the first transition temperature $T_{xy}$, we have determined from visualizing the simulation results that the adsorbate begins migrating within the plane of the GNMR.  During the in-plane migration, the adsorbate migrates on top of the GNMR, but remains connected to the GNMR through the van der Waals interactions.  A direct result of the in-plane migration is the excitation of many in-plane vibrational modes in the GNMR that result from the friction between the migrating adsorbate and the GNMR.  As a result, the Q-factor of the GNMR is substantially reduced at $T_{xy}$.  We note that the Q-factor at $T<7$~{K} is so high (i.e. that energy dissipation is quite weak, as has been observed experimentally\cite{BunchJS2007sci,EichlerA,ZandeAMVD,ChenC2009nn}) that it is difficult for us to precisely extract its value from our MD simulations.  As a result, we assign the same Q-factor of 30000 at these ultra-low temperatures, which is likely to be a lower bound estimate.  Hence, experimentalists should observe a larger decrease in the inverse of the Q-factor at $T_{xy}$ than is predicted by the MD simulations.

\begin{figure}[htpb]
  \begin{center}
    \scalebox{0.9}[0.9]{\includegraphics[width=8cm]{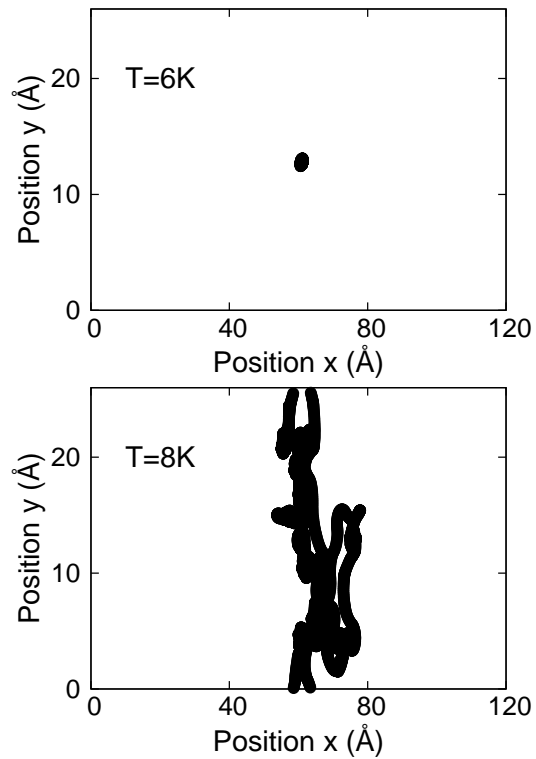}}
  \end{center}
  \caption{(Color Online) The trajectory of the adsorbate in (a), where $T<T_{xy}$ shows a localized vibration of this adsorbate at 6~{K}.  In contrast, figure (b) where $T>T_{xy}$ shows a migration of the adsorbate at 8~{K}. The migration temperature is determined to be $T_{xy}=7\pm 1$~{K}.}
  \label{fig_trajectory_xy}
\end{figure}

\begin{figure}[htpb]
  \begin{center}
    \scalebox{0.9}[0.9]{\includegraphics[width=8cm]{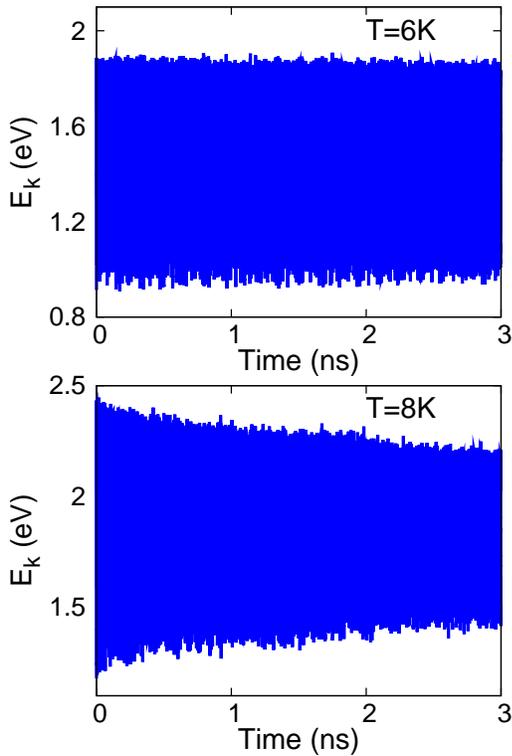}}
  \end{center}
  \caption{(Color Online) The kinetic energy time history for a GNMR of size (50, 6) with a single adsorbate. The energy dissipation is obviously higher at 8~{K} than at 6~{K}, which demonstrates the strong dissipative effects that emerge from the in-plane adsorbate migration on the Q-factor of the GNMR.}
  \label{fig_energy_xy}
\end{figure}

\begin{figure}[htpb]
  \begin{center}
    \scalebox{1.0}[1.0]{\includegraphics[width=8cm]{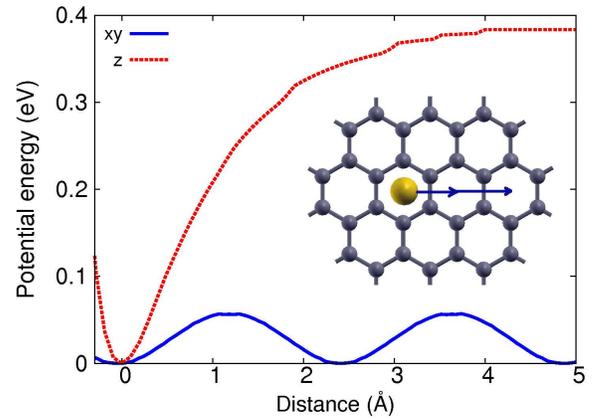}}
  \end{center}
  \caption{(Color Online) The potential energy barrier corresponding to the adsorbate migration in the $xy$ (in)-plane (blue solid line) or $z$ (out-of-plane) direction (red dashed line). Inset shows the $xy$ migration direction.}
  \label{fig_potential_diffusion}
\end{figure}

Figure.~\ref{fig_trajectory_xy} shows the $xy$ trajectory of the adsorbate at temperatures 6 and 8~{K}, which is just below or above the first transition temperature $T_{xy}=7$ K. Panel (a) shows that the motion of the adsorbate at 6~{K} is very localized.  In contrast, panel (b) shows that at 8~{K}, the adsorbate migrates around the surface of the GNMR.  The kinetic energy time history at these two temperatures are compared in Fig.~\ref{fig_energy_xy}, where the energy dissipation is obviously higher at 8~{K} than that of 6~{K}, showing how the adsorbate migration strongly degrades the Q-factors of GNMRs.  These results provide clear evidence that the mechanism controlling the step-like jump in the Q-factor in Fig.~\ref{fig_qfactor} at $T_{xy}=7$ K is the initiation of the in-plane adsorbate migration.

At the second transition at $T_{z}$ in Fig.~\ref{fig_qfactor}, the inverse of the Q-factor scales as $T^{\alpha}$ for $T<T_{z}$ with $\alpha=0.29$. For $T>T_{z}$, the inverse of the Q-factor scales as $T^{\beta}$, with $\beta =0.87$. This transition is exactly the temperature scaling of the Q-factor corresponds to that which has been observed experimentally, where the corresponding scaling factors are $\alpha=0.35$ and $\beta=2.3$ in the experiment.\cite{ZandeAMVD} There are several possible effects that may be responsible for this difference in the scaling exponents between our MD simulations and the experiments, as will be shown in the following. We observe from visualizing the simulation results that the critical temperature $T_{z}$ is exactly the temperature at which the adsorbate begins to migrate out of the plane, i.e. in the $z$-direction. During this migration, the adsorbate develops sufficient kinetic energy such that it breaks the van der Waals bond connecting it to the GNMR, and travels away from the GNMR surface.  Due to the finite size of the simulation box, the migrating adsorbate eventually reaches and reflects from the boundary of the simulation box, which results in it hitting the GNMR frequently from arbitrary directions.  These arbitrary collisions cause the excitation of various vibrational modes, including both in-plane and out-of-plane modes.  If the incident angle of the adsorbate is orthogonal to the vibrational motion of the GNMR, then the out-of-plane modes will predominately be excited, which strongly impacts the out-of-plane mechanical vibrations of the GNMR.  As a result, the value in the MD simulations of $\beta$ are always larger than $\alpha$.

This point can also be understood from an energetic point of view. Fig.~\ref{fig_potential_diffusion} shows that substantially different potential energy barriers exist for adsorbate migration in the in-plane ($xy$) as compared to the out-of-plane ($z$) directions.  Specifically, the energetic barrier is substantially smaller for the $xy$ in-plane migration than the $z$-direction out-of-plane migration.  This is because the adsorbate will be attracted by neighboring carbon atoms during its migration process from one minimum potential energy position to a neighboring minimum potential energy position. As a result, the energy exchange between the adsorbate and the GNMR is larger for the $z$ migration than the $xy$ migration. In other words, the $z$ migration has a stronger influence on the GNMR oscillation.

The second transition at $T_{z}$ is continuous, which is in agreement with experiments.\cite{ZandeAMVD,ChenC2009nn} It is different from the step-like jump in the Q-factor at $T_{xy}$ because the migration of the adsorbate changes gradually from the in-plane migration into the out-of-plane migration.

\begin{figure}[htpb]
  \begin{center}
    \scalebox{0.9}[0.9]{\includegraphics[width=8cm]{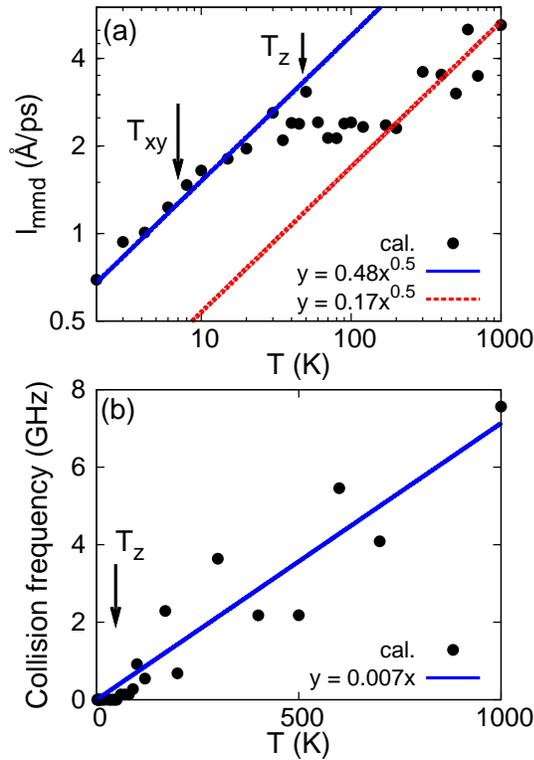}}
  \end{center}
  \caption{(Color Online) MMD of the adsorbate in a GNMR of size (50, 6) with a single adsorbate. (a) shows the MMD. The obvious transition at $T_{z}$ is with respect to the out-of-plane migration of the adsorbate. (b) shows the frequency of the collision between the adsorbate and the GNMR. The linear temperature dependence at high temperature reflects the free motion of the adsorbate after the out-of-plane migration. The two arrows indicate the two transition temperatures $T_{\rm xy}$ and $T_{z}$.}
  \label{fig_mfp}
\end{figure}

\begin{figure}[htpb]
  \begin{center}
    \scalebox{0.9}[0.9]{\includegraphics[width=8cm]{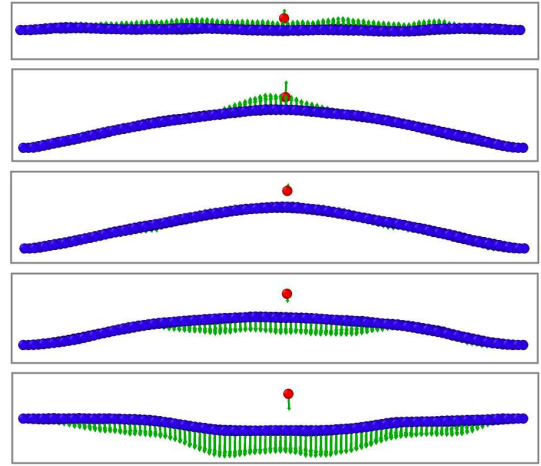}}
  \end{center}
  \caption{(Color Online) An unsuccessful escape attempt of the adsorbate from the GNMR at 50~{K}. From top to bottom, the adsorbate is trying to escape from the GNMR surface, but is unsuccessful because the energy imparted to it from the oscillation of the GNMR is not large enough to break the van der Waals bonds connecting it to the GNMR. Velocity vectors are represented by arrrows.}
  \label{fig_noescape_movie_50K}
\end{figure}

\begin{figure}[htpb]
  \begin{center}
    \scalebox{0.9}[0.9]{\includegraphics[width=8cm]{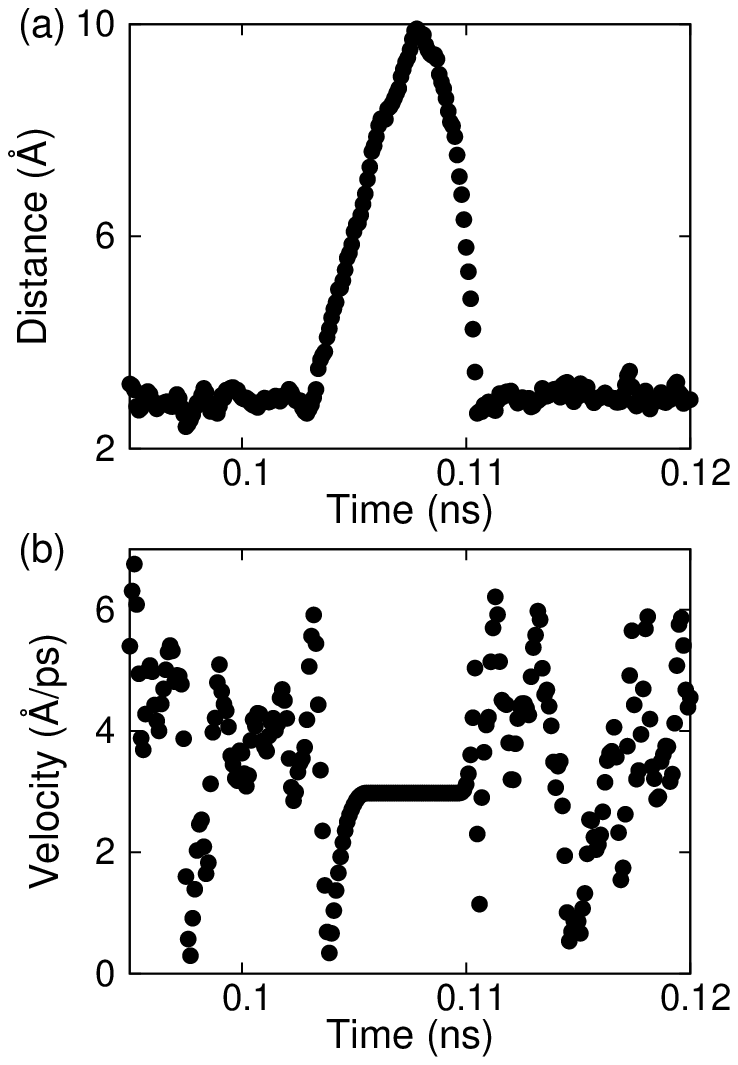}}
  \end{center}
  \caption{(Color Online) The distance and velocity of the adsorbate during an unsuccessful escape attempt. (a) shows the nearest distance between the adsorbate and the GNMR. (b) shows the velocity of the adsorbate during the escaping process.}
  \label{fig_noescape_50K}
\end{figure}

\begin{figure}[htpb]
  \begin{center}
    \scalebox{0.9}[0.9]{\includegraphics[width=8cm]{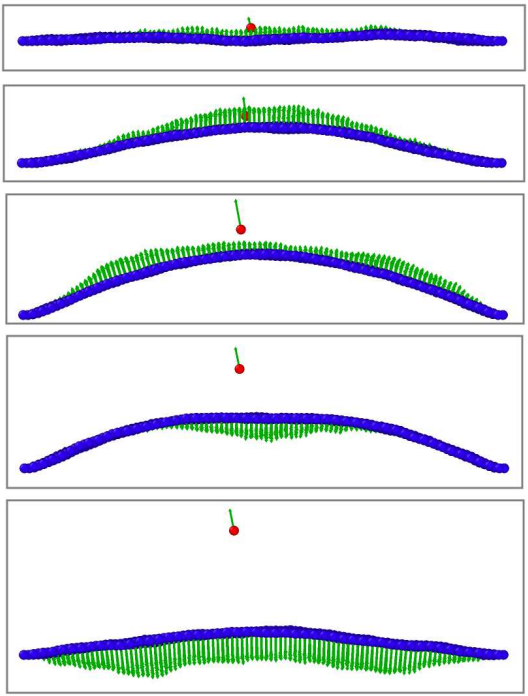}}
  \end{center}
  \caption{(Color Online) A successful escape of the adsorbate from the GNMR at 300~{K}. From top to bottom, the adsorbate succeeds in escaping from the GNMR surface.  Velocity vectors are represented by arrrows.}
  \label{fig_escape_movie_300K}
\end{figure}

\begin{figure}[htpb]
  \begin{center}
    \scalebox{0.9}[0.9]{\includegraphics[width=8cm]{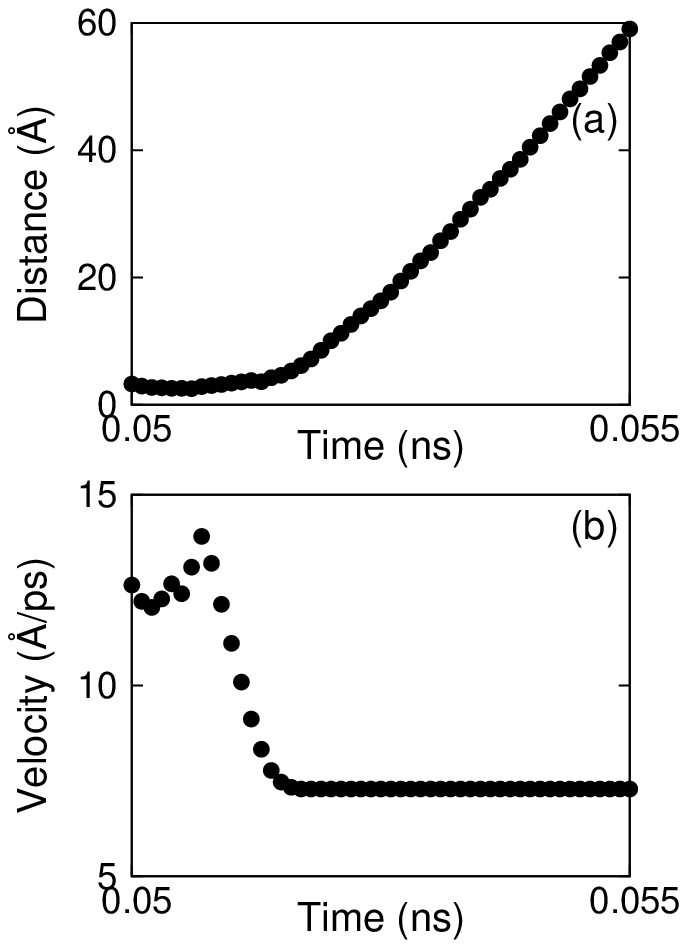}}
  \end{center}
  \caption{(Color Online) The distance and velocity of the adsorbate during a successful escape. (a) shows the nearest distance between the adsorbate and the GNMR. (b) shows the velocity of the adsorbate during the escaping process.}
  \label{fig_escape_300K}
\end{figure}

\begin{figure}[htpb]
  \begin{center}
    \scalebox{1.0}[1.0]{\includegraphics[width=8cm]{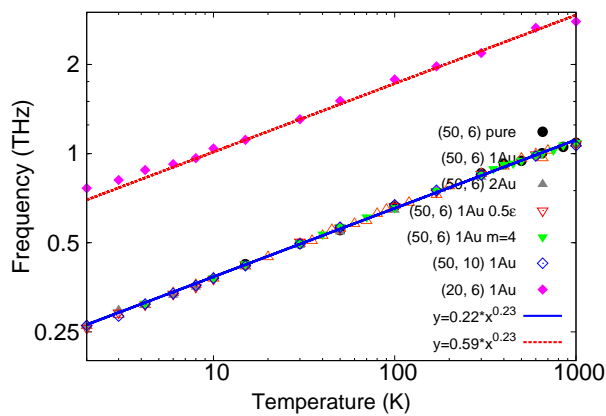}}
  \end{center}
  \caption{(Color Online) The frequency of the GNMR with different size and various adsorbate coverage conditions. The length ($L$) is important, as the frequency of the oscillation mode is proportional to $1/L^{2}$ due to the flexural properties of the two-dimensional graphene sheet.}
  \label{fig_f_all}
\end{figure}

\begin{figure*}[htpb]
  \begin{center}
    \scalebox{0.9}[0.9]{\includegraphics[width=\textwidth]{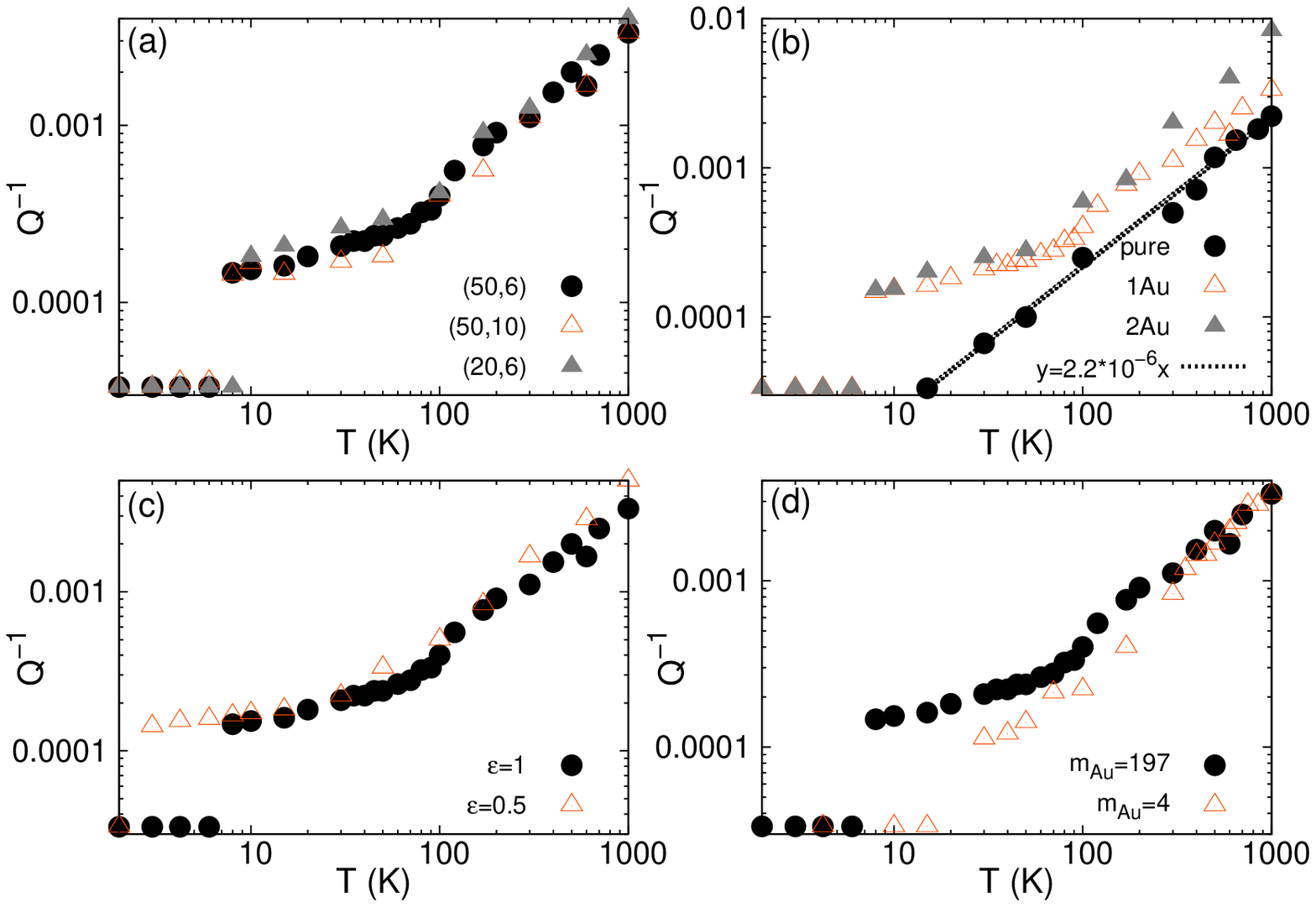}}
  \end{center}
  \caption{(Color Online) Various effects on the temperature scaling in the inverse of the Q-factor. (a) shows similar temperature scaling in GNMRs of different dimensions when a single adsorbate is considered. (b) The temperature scaling in GNMRs with different numbers of adsorbate. (c) The temperature scaling for GNMRs with different interaction strength between the adsorbate and the GNMR. $\epsilon=1$ corresponds to the actual LJ energy parameter (0.02936~{eV}) between Au and C. (d) The temperature scaling in the case of different adsorbate masses.}
  \label{fig_qfactor_all}
\end{figure*}

To investigate the temperature scaling further, we calculate the mean migration distance (MMD) ($l_{\rm mmd}$) of the adsorbate, where the MMD is defined to be the average distance of the adsorbate per unit time. It is calculated by:
\begin{eqnarray}
l_{\rm mmd} = \frac{1}{(N-1)\Delta t} \sum_{i=1}^{N-1} |\vec{r}(t_{i+1})-\vec{r}(t_{i})|,
\end{eqnarray}
where $N$ is the total MD simulation steps, $\vec{r}(t)$ is the trajectory of the adsorbate from the MD simulation, and $\Delta t$ is the time step in the MD simulation. 

The MMD is shown in Fig.~\ref{fig_mfp}~(a) for a single adsorbate on top of the GNMR with size (50, 6). The MMD is proportional to $\sqrt{T}$ in the whole temperature range except $T\in[47.5, 200.0]$~K. For $T<T_{z}$, the MMD is proportional to $\sqrt{T}$. While no obvious transition is observed in the MMD at $T_{xy}$, an obvious transition occurs at $T_{z}$.  Interestingly, in the high temperature region, the MMD is also proportional to $\sqrt{T}$. These results are analyzed as follows. For $T<T_{xy}$, the adsorbate migration is minimal, and because its vibrational amplitude is proportional to $\sqrt{T}$, the MMD is proportional to $\sqrt{T}$. For temperatures $T_{xy}<T<T_{z}$, the adsorbate migrates in the plane on top of the GNMR surface, and so the MMD is proportional to its $xy$ velocities, which depends on temperature as $\sqrt{T}$. Hence, the MMD within this temperature range also increases as $\sqrt{T}$. At temperatures above $T_{z}$, the out-of-plane migration happens, so the adsorbate moves freely without being attracted to the GNMR surface, and as a result, the MMD is proportional to its velocity, which depends on temperature as $\sqrt{T}$.  

Some disturbance can be found in the MMD at $T_{z}$ in Figure.~\ref{fig_mfp}(a), which indicates that the adsorbate has begun the out-of-plane migration. In this stage, the adsorbate is trying to break the van der Waals bonds connecting it to the GNMR, but fails as the kinetic energy it gains from the GNMR is not sufficient to break the van der Waals bond.  

Thus, in this temperature range, [47.5, 200.0]~K, the adsorbate loses a substantial amount of its velocity whenever it attempts to escape, which accounts for the disturbance in the MMD seen at $T_{z}$.  A specific example of this, i.e. an unsuccessful escape attempt at 50~K, is shown in Figure.~\ref{fig_noescape_movie_50K}. From top to bottom, the adsorbate attempts, but eventually fails to escape and thus is pulled back towards the GNMR surface as the kinetic energy it gains from the mechanical oscillation of the GNMR is not large enough to break its van der Waals with the GNMR. Fig.~\ref{fig_noescape_50K}~(a) shows the distance between the adsorbate and the GNMR during this unsuccessful escape, which illustrates that the adsorbate is dragged back by the GNMR after the escape attempt. Panel (b) shows the corresponding velocity of the adsorbate. During the whole unsuccessful escape attempt, the velocity of the adsorbate is generally lower than the value required by the equipartition theorem. As a result, this unsuccessful escape leads to the disturbance in the MMD seen in Figure.~\ref{fig_mfp}(a).  

With increasing temperature, more and more escape attempts become successful, because the GNMR can transfer sufficient kinetic energy to the adsorbate to allow it to escape from the GNMR surface.  Fig.~\ref{fig_escape_movie_300K} shows a successful escape for the adsorbate at 300~{K}. The kinetic energy passed to the adsorbate is large enough that the velocity of the adsorbate points in the outward direction with respect to the GNMR at the moment when the GNMR has zero velocity and maximum potential. Fig.~\ref{fig_escape_300K} shows the corresponding distance and the velocity of the adsorbate, which displays clearly the uniform motion of the adsorbate after breaking the van der Waals bond connecting it to the GNMR.

Figure.~\ref{fig_mfp}~(b) shows the collision frequency between the adsorbate and the GNMR. The collisions occur after the emergence of the out-of-plane migration for the adsorbate at temperatures above $T_{z}$.  As previously discussed, the adsorbate eventually rebounds from the boundary of the simulation cell and thus collides frequently with the oscillating GNMR. The collision frequency is calculated by the total collision counts divided by the total MD simulation time. In the high temperature range, the collision frequency increases linearly with increasing temperature, because both the free migrating velocity of the adsorbate and the oscillation velocity of the GNMR are proportional to $\sqrt{T}$. The combination of these two $\sqrt{T}$ scaling relations gives the linear temperature dependence seen in Figure.~\ref{fig_mfp}(b).

Fig.~\ref{fig_f_all} shows the frequency of the GNMRs with different sizes and doped by adsorbates with different mass ($m_{\rm Au}$) or the LJ energy parameter ($\epsilon$). The length $l_{x}$ is important, as the frequency of the bending mode is proportional to $1/l_{x}^{2}$. The other effects from different adsorbate mass or LJ energy parameter are not obvious as compared with the temperature effect in the figure. In all GNMRs, the frequency increases with increasing temperature following a power function with power factor 0.23 because of the vibration-induced effective strain within the GNMR.\cite{JiangJW2012nanotechnology}

Finally, Fig.~\ref{fig_qfactor_all} shows how various effects impact the temperature scaling in the Q-factor of GNMRs. Panel (a) shows that changing the size of the GNMR only slightly modifies the value of the two transition temperatures $T_{xy}$ and $T_{z}$, as well as the two scaling factors $\alpha$ and $\beta$. Hence the size effect is not very important in the temperature scaling phenomenon of the Q-factor, so this phenomenon should be observed in GNMR samples of all sizes. 

Panel (b) shows how increasing the number of adsorbates impacts the temperature scaling. For pure GNMRs without adsorbates, the Q-factor shows a linear temperature dependence. The power factor $\alpha$ is almost the same in both situations; while the power factor $\beta$ are 0.87 and 1.27 in GNMR with one adsorbate (Au$_{1}$) and two adsorbates (Au$_{2}$), respectively. This significant change ($46\%$ difference in $\beta$) displays that the number of adsorbates has a substantial effect on the temperature scaling, where in particular the transition at $T_{z}$ becomes sharper in GNMRs with more adsorbates. We note that this effect should be experimentally verifiable.

Panel (c) shows that the strength of the van der Waals interaction between the adsorbate and the GNMR has an important effect on the value of the two critical temperatures $T_{xy}$ and $T_{z}$.  Specifically, those critical temperatures are reduced by nearly 50\% if the interaction parameter $\epsilon$ between adsorbate and the GNMR is reduced by 50\%, while $\alpha$ and $\beta$ are not obviously affected.  

Finally, panel (d) shows that the mass of the adsorbate also plays an important role for the temperature scaling. For lighter adsorbates, the temperature scaling is smoother at $T_{z}$. Both transition temperatures $T_{xy}$ and $T_{z}$ become much higher for the lighter adsorbate, since those atoms have a smaller kinetic energy at the same temperature as larger mass adsorbates, and therefore more thermal energy is needed to induce migration for the smaller mass atoms. Panels (c) and (d) may serve as useful guidelines for experimentalists to verify our theoretical predictions, as different types of adsorbates (of different mass or/and different interaction strengths with GNMRs) are usually observed experimentally.

\section{conclusion}
In conclusion, we have performed classical MD simulations to study how adsorbate migration affects the Q-factors of GNMRs.  Our key finding is that there are two critical temperatures, which correspond to the initiation of in-plane and out-of-plane migration of the adsorbate.  The out-of-plane adsorbate migration is found to be the origin for the temperature scaling in the Q-factor observed experimentally\cite{ZandeAMVD,ChenC2009nn}, because the resonant oscillation of the GNMR is strongly interrupted by collisions from the migrating adsorbate. We predict a discontinuous decrease of the Q-factor at the lower critical temperature due to the initiation of in-plane migration of the adsorbate. We also studied, in the hopes of providing useful guidelines to experimentalists to verify these theoretical predictions, how different numbers of adsorbate atoms, sizes of the GNMR, or strength of van der Waals interactions between the adsorbate and the GNMR impact the temperature scaling of the Q-factor.

\textbf{Acknowledgements} JWJ would like to point out that the idea of the out-of-plane migration is stimulated by a game with his first baby CC, who is crazy about putting everything on top of the bed sheet and vibrating the sheet to throw things into the air. The work is supported by the German Research Foundation (DFG).  HSP acknowledges the support of NSF-CMMI 1036460. All authors thank Andreas Isacsson for valuable discussions in the initial stage of the work.

%\bibliographystyle{aipnum4-1}
%\bibliography{biball}

\end{document}